\documentclass[aps,preprint,prl,showpacs,superscriptaddress,groupedaddress]{revtex4-1} 

\usepackage{verbatim}
\usepackage{multirow}
\usepackage{dcolumn}   % needed for some tables
\usepackage{bm}        % for math
\usepackage{amssymb}   % for math
\usepackage{amsmath}
\usepackage{amsfonts}
\usepackage{graphicx}
\usepackage[ngerman, english]{babel}
\usepackage{float}
\usepackage{youngtab}
\usepackage[utf8]{inputenc}
\usepackage{mathrsfs}
\usepackage{rotating}
\usepackage{braket}
\usepackage{ulem}
\usepackage{dsfont}
\usepackage{subfigure}
\usepackage{color}
\usepackage{nicefrac}
\usepackage{slashed}
\usepackage{hyperref}

\newcommand{\halb}{{\nicefrac{1}{2}}}

\begin{document}
%\preprint{-Preprint number-}
\title{Sterile Neutrino Shortcuts in Asymmetrically Warped Extra Dimensions}
\author{Dominik D\"oring}
\email[corresponding author: ]{dominik.doering@tu-dortmund.de}
\author{Heinrich P\"as}
\email[]{heinrich.paes@tu-dortmund.de}
\affiliation{Fakult\"at f\"ur Physik,
        Technische Universit\"at Dortmund, 44221 Dortmund, Germany}
\preprint{DO-TH 18/20}

\begin{abstract}
Light sterile neutrinos are a popular extension of the Standard Model and are being discussed as a possible
explanation for various neutrino oscillation anomalies, including the LSND, MiniBooNE, Reactor and Gallium anomalies. 
In order to avoid inconsistencies with constraints derived from disappearance experiments and cosmology, altered dispersion relations
- which may originate from extra dimensions - 
have been proposed as a possible solution, dubbed as "neutrino shortcuts
in the extra dimension".
In this paper we develop a neutrino mass model with 
an asymmetrically warped extra dimension and two additional gauge singlet neutrinos, 
one being responsible for neutrino mass generation, 
while the other one is allowed to propagate in the extra dimension, giving rise to the desired change of the dispersion relation on the brane.
By compactifying the extra-dimensional theory on an $S^1/\mathbb{Z}_2$ orbifold,
deriving the shape of the Kaluza-Klein tower and identifying the effective sterile neutrino dispersion relation on the brane, 
we can demonstrate that the earlier, phenomenological models are recovered as the 4-dimensional effective field theory limit
of the model discussed here.
 
\end{abstract}

\maketitle
\section{Introduction}
Sterile neutrinos are a common prediction in many neutrino mass models and have been proposed as a possible 
solution for various neutrino anomalies, hints for inconsistencies in cosmological data, and as a possible dark matter candidate. In particular sterile neutrinos with masses in 1~eV mass range are discussed in the context of
the LSND, MiniBooNE, Reactor and Gallium anomalies. There exist, however, stringent constraints on light sterile
neutrinos, both from neutrino oscillation experiments as well as from cosmology, which rule out the most simple scenarios. A possible way out of this dilemma is the hypothesis that sterile
neutrinos may feature effective Lorentz violating corrections to the standard dispersion relation 
$E^2 = p^2 + m^2$, which leads to an interesting and rich phenomenology. A particularly attractive  
realization giving rise to such altered dispersion relations (ADRs) are scenarios where the sterile neutrinos
can take shortcuts in extra dimensions. It has been conjectured in the past that this phenomenon arises naturally 
in models where the sterile neutrino propagates in an asymmetrically warped spacetime
\cite{PaesShortcut,PaesHollenbergBaseline,Aeikens:2016rep,Aeikens:2014yga,SuperluminalNeutrinos,Doring:2018cob}.

Models with large extra dimensions became popular in the in the late 1990s,
when it was discovered that the hierarchy problem could be resolved or ameliorated by
adopting several flat extra dimensions \cite{ADD} or one 
compactified, warped extra dimension  as in  the well-known 'Randall-Sundrum 1 (RS1)' model \cite{RS}. 
In such theories, typically the SM particle content is located on the 3-brane, while gauge singlets (like the graviton or sterile neutrinos) are allowed to propagate in the extra dimension and therefore experience the associated
warping. 

While the RS model uses symmetrically warped spacetime of the form
 $\mathrm{d}s^2 = e^{-2kr\phi} \eta_{\mu \nu} \mathrm{d}x^\mu
 \mathrm{d}x^\nu + r^2 \mathrm{d}\phi^2$, asymmetrically warped metrics
 with the form
 $\mathrm{d}s^2 = -A^2(\phi) \mathrm{d}t^2+  B^2(\phi) \mathrm{d}\vec{x}^2
 + C^2(\phi) r^2 \mathrm{d}\phi^2$ arise from simple bulk sources and are
 linked to the cosmological constant problem \cite{Csaki:2000dm} and
 the horizon problem \cite{Chung:1999xg,Chung:1999zs}.
 Whereas these kind of models preserve Lorentz symmetry in the 5D theory,
 they predict 4D Lorentz violation on the brane via an altered dispersion
 relation for sterile neutrinos,
 which proves to be helpful for the solution of the current anomalies in
 short baseline and reactor neutrino oscillation data.
 Scenarios with altered dispersion relations adopt additional terms in the
 usual relation between energy $E$ and momentum $\vec{p}$,
 $E^2 \neq \left|\vec{p}\right|^2 + m^2$.
 Energy dependent elements of the mixing matrix and mass squared
 differences can be generated by an additional effective potential in the
 Hamiltonian in flavor space and thus
 may pose an explanation for the anomalies encountered in short-baseline
neutrino oscillation data such as the LSND (\cite{LSND1,LSND2,LSND3,LSND4,LSND5}) or MiniBooNE (\cite{MiniBooNECombined,MiniBooNE2}) anomalies. Quite recently the MiniBooNE collaboration has
reported an evidence of $4.8\sigma$ for new physics beyond the Standard Model which, combined with the LSND experiment, increases to $6.1~\sigma$ \cite{MiniBooNE2018}. New efforts to clarify this situation are planned or under development  \cite{ThreeDetectorSBNO,TeraoMicroBooNE}). 
In this paper we thus develop a neutrino mass model featuring an asymmetrically-warped extra dimension
which justifies the effective 4-dimensional low-energy "sterile neutrino shortcut" phenomenology
proposed in \cite{PaesShortcut,PaesHollenbergBaseline,Aeikens:2016rep,Aeikens:2014yga,SuperluminalNeutrinos,Doring:2018cob}. 

\section{An asymmetrically warped neutrino mass model}
 
Oribolding the fifth dimension on $S^1/\mathbb{Z}_2$ allows to 
parameterize the extra dimension by an angular coordinate $\phi$ and an extra 
dimensional radius $r$. It also ensures that the $\phi$-coordinate satisfies the periodic boundary conditions $\phi = \phi + 2\pi$ and $Z_2$-symmetry $\phi=-\phi$. Hence, the extra dimension can be
entirely described with values for $\phi$ in the range $0 \leq \phi \leq \pi$. Just like in the RS-model, orbifold fixed points will be populated by $3$-branes, corresponding to standard (3+1) Minkowskian spactime parametrized by the 
coordinates $x^\mu$. For a sufficiently general ansatz we use a metric tensor $G_{MN}$ of the form
\begin{align}
 \label{eq:MetricTensor}
 \left(G_{MN}\right) = 
  \begin{pmatrix}
    -A^2 & & & & \\
    & B^2 & & & \\
    & & B^2 & & \\
    & & & B^2 & \\
    & & & & (rC)^2 \\ 
  \end{pmatrix} \, ,
\end{align}
where where the metric elements $A=A(\phi)$, $B=B(\phi)$, $C=C(\phi)$ are functions of the extra dimensional angular coordinate $\phi$ 
and the latin indices $M,N = 0,1,2,3,4$ imply a five-dimensional metric, and are chosen to recover Minkowskian space $\eta_{\mu \nu}$ on the $3$-branes.
This class of non-factorizable spacetimes is called 'asymmetrically warped'. 
Note that such a metric tensor does not represent a vacuum solution of Einstein's equations, but can be achieved e.g. by introducing simple bulk sources \cite{Csaki:2000dm}. 
We adopt
a single fermionic SM singlet field $\Psi$ to be allowed to enter the extra dimension, thus the general action for such a Dirac fermion is \cite{NeubertGrossman}
\begin{align}
\label{eq:mostgeneralaction}
  S = \int \mathrm{d}^4 x \int \mathrm{d}\phi \sqrt{\det{G}} \left\lbrace E_a^A \left[ 
  \frac{i}{2} \overline{\Psi} \gamma^a \left( \partial_A - \overleftarrow{\partial_A } \right)\Psi
  + \frac{\omega_{bcA}}{8} \overline{\Psi} \left\lbrace \gamma^a , \sigma^{bc} \right\rbrace \Psi \right] - m \,  \text{sgn}(\phi)
  \overline{\Psi} \Psi
  \right\rbrace \, ,
\end{align}
where $E^A_a$ denotes the inverse Vielbein, $\omega_{bcA}$ is the spin connection, $\sigma^{bc}=[\gamma^b, \gamma^c]$ is the commutator of the Dirac matrices and $m$ is the fermion's fundamental Dirac mass.
The Vielbein is defined via $G^{MN} = \eta^{mn} E_m^M E_n^N$
as the transformation of a coordinate basis of basis vectors $\partial_a$ into another, equivalent basis $e_A = E^a_A \partial_a$. This allows for a conversion of spacetime indices $A$ to Lorentz indices $a$ in the local
tangent space. For this to be true, the Vielbein has to be non-singular. For the metric~\ref{eq:MetricTensor} we obtain $\left(E_a^A\right) = \text{diag}\left( \frac{1}{A}, \frac{1}{B},\frac{1}{B},\frac{1}{B},\frac{1}{rC}\right)$. \\
The gamma-matrices $\gamma^a$ obey the Clifford-Algebra $\left\lbrace \gamma^a, \gamma^b \right\rbrace = 2\eta^{ab}$ ,
where $\eta^{ab}$ is the Minkowski-metric, with $\gamma^{a=4} = i\gamma^5$ . The mass term of Eq. \eqref{eq:mostgeneralaction} contains the sign of the extra dimensional coordinate in order to preserve $Z_2$-symmetry. 
It can be shown that the spin connection term $ E_a^A \frac{\omega_{bcA}}{8} \overline{\Psi} \left\lbrace \gamma^a , \sigma^{bc} \right\rbrace \Psi$ vanishes so that we are left with
\begin{align}
\label{eq:generalaction}
  S = \int \mathrm{d}^4 x \int \mathrm{d}\phi \sqrt{\det{G}} \left\lbrace E_a^A \left[ 
  \frac{i}{2} \overline{\Psi} \gamma^a \left( \partial_A - \overleftarrow{\partial_A } \right)\Psi \right] - m \,  \text{sgn}(\phi)
  \overline{\Psi} \Psi
  \right\rbrace \, .
\end{align}

%\section{Kaluza-Klein spectrum}

We now decompose the Dirac spinor $\Psi$ using the chiral operator $\Psi_\halb = \frac{1 \mp \gamma^5}{2} \Psi$. Note that we do not explicitly call these spinors left- or right-handed, since in five dimensions this
concept cannot be applied. This is because $\gamma^5$ is part of the Clifford Algebra in 5D and therefore cannot serve as a chiral projector. This is true for any odd dimensional spacetime. \\ 
After an integration by parts we obtain expressions for the action, which can be associated with a mass-term and a kinetic-term respectively, because of their spinor structure (for details see appendix A).
The expressions yield
\begin{align}
\label{eq:kineticaction}
S_{\text{kin}} = \int \mathrm{d}^4 x \int \mathrm{d}\phi \sqrt{\text{det}G} \left\lbrace 
\overline{\Psi}_1 \, i \left[ \frac{\gamma^0}{A} \partial_0 + \frac{\gamma^k}{B} \partial_k \right] \Psi_1 
+\overline{\Psi}_2 \, i \left[ \frac{\gamma^0}{A} \partial_0 + \frac{\gamma^k}{B} \partial_k \right] \Psi_2 \right\rbrace \, 
\end{align}
and 
 \begin{align}
 \begin{split}
 \label{eq:massaction}
S_{\text{mass}} = \int \mathrm{d}^4 x \int \mathrm{d}\phi  \left\lbrace \vphantom{\frac{\sqrt{\text{det}G}}{2rC}} \right. 
- \left[\vphantom{\frac{\sqrt{\text{det}G}}{2rC}} \right.  &\overline{\Psi}_1 \left( \frac{\sqrt{\text{det}G}}{2rC} \, \partial_\phi + \partial_\phi \frac{\sqrt{\text{det}G}}{2rC} \, 
\right) \Psi_2 \\
  - &\overline{\Psi}_2 \left( \frac{\sqrt{\text{det}G}}{2rC} \, \partial_\phi + \partial_\phi \frac{\sqrt{\text{det}G}}{2rC} \, 
\right) \Psi_1 \left. \vphantom{\frac{\sqrt{\text{det}G}}{2rC}} \right] \\
&-m \, \text{sgn}(\phi) \left[ \, \overline{\Psi}_1 \Psi_2 + \overline{\Psi}_2 \Psi_1 \right] 
\left.\vphantom{\frac{\sqrt{\text{det}G}}{2rC}} \right\rbrace  \, .
 \end{split}
\end{align}

We apply a Kaluza-Klein (KK) decomposition, i.e. we expand the 5D spinors  $\Psi_\halb(x, \phi)$ in a in a series of a product of functions $ \psi^{\halb}_n(x)$ and $\hat{f}^\halb_n(\phi)$
\begin{align}
\label{eq:ansatz}
 \Psi_{\halb} = \sum_n \psi^{\halb}_n(x) \frac{1}{\sqrt{2r\xi}} \hat{f}^\halb_n(\phi) \hspace{2cm} 
 \text{with} \hspace{0.5cm} \xi = \xi(\phi)= \frac{\sqrt{\text{det}G}}{2rC} \, ,
\end{align}
where $\hat{f}^\halb_n(\phi)$ will be constructed as eigenfunctions of a Hermitian operator. This operator arises as we compare the decomposed action to the standard Dirac action in 4D.
It can be shown that the kinetic part of the action is actually able to recover the 4D Dirac actions kinetic part 
\begin{align}
  \label{eq:correctedDirac}
 S_{\text{DiracKin}} = \int \mathrm{d}^4 x \left\lbrace \overline{\psi_n}(x) \left( i \slashed \partial +  \hat{\Omega} \right) \psi_n(x)
 \right\rbrace 
\end{align}
up to some correction $\Omega$ by choosing the scalar product 
\begin{align}
\label{eq:scalarproduct}
 \int \mathrm{d}\phi \hat{f}_n^\halb \frac{C}{A} \hat{f}_m^{\halb \dagger} := \delta_{nm} \, .
\end{align}
for the functions $\hat{f}^\halb_n(\phi)$. The correction term can be identified as 
\begin{align}
  \label{eq:OmegaTerm}
 \Omega = \int \mathrm{d}\phi \sum_{mn} \sum^2_{j=1}\left[ \overline{\psi^j_n} \hat{f}_n^{j \dagger} 
\frac{C(A-B)}{AB} \, i \, \gamma^k \partial_k \,  \psi^j_m \hat{f}_m^{j} \right] \, .
\end{align}
This term contains an operator, which induces Lorentz violation (LV) in the 4D projection, while being Lorentz invariant in the full 5D picture. This can lead to a different interplay between the momentum and the 
energy of a particle on the brane (see e.g. \cite{KosteleckyShortBaseline,KosteleckyNeutrinos,Antonelli}). In other words, the operator changes the
dispersion relation $E^2 = \vec{p}^2 + m^2$ experienced by an observer on the brane.
Applying Eq. \eqref{eq:scalarproduct} to the decomposed $S_{\text{mass}}$ and matching it to the standard mass term of the Dirac action, we can derive another condition
\begin{align}
\label{eq:dglasymm}
 \left( \mp \frac{\partial_\phi}{r} - m C \right) \hat{f}_k^{\nicefrac{2}{1}} = - M_k \frac{C}{A} \hat{f}_k^\halb \, .
\end{align}
for the extra dimensional function $\hat{f}_k^{\halb}$. This is a system of first order, coupled, eigenvalue-like equations, which determine the behavior of the extra dimensional function $\hat{f}_\halb$ and therefore the
shape of the KK spectrum of masses. From this expression we can derive that the shape of the KK tower in the asymmetrically warped case does not differ from symmetric warping scenarios 
($\mathrm{d}s^2= F(\phi)\eta_{\mu \nu} \mathrm{d}x^\mu \mathrm{d}x^\nu + r^2 \mathrm{d}\phi $). This is due to the non-dependence on the metric element $B(\phi)$. In an RS-like set-up, the KK spectrum is shaped like the 
roots of Bessel's function. The order of this function is determined by the extra dimensional fermion's fundamental mass and the inverse radius of the extra dimension (for further reading see \cite{NeubertGrossman}).
The only difference between symmetric and asymmetric scenarios is therefore the
induced LV on the brane, which is dependent on the difference $A(\phi)-B(\phi)$. Obviously, the LV vanishes in the symmetric limit, recovering the results of \cite{NeubertGrossman}.

\section{Altered Dispersion Relation and Connections to the Shortcut Parameter}
To study the LV on the brane quantitatively, we extract the aforementioned altered dispersion relation (ADR) on the brane from the correction term (Eq. \eqref{eq:OmegaTerm}) of the underlying action $S$.
By using Eq. \eqref{eq:scalarproduct}, we can express this correction term as 
\begin{align}
S \supset \int \mathrm{d}^4 x 
\sum_{n,m} \sum^2_{j=1} \left[\, \overline{\psi_n^j} \, \tilde{I}^j_{nm} \, i \, \gamma^k \partial_k \,  \psi^j_m  
\right] ,
\end{align}
where $\tilde{I}_{nm} = \int \mathrm{d}\phi \hat{f}^{j \dagger}_n \frac{C}{B} \hat{f}^{j}_m - \delta_{nm}$ is the correction parameter.
To study neutrino oscillation properties, we now introduce an active, brane-bound,
lefthanded neutrino state $\nu_L$ to the action and consider only the lefthanded zero-mode $\Psi_L^0$ of the extra dimensional singlet state, without taking into account the Kaluza-Klein excitations (this can be justified by
adopting a sufficiently small extra dimension).
A righthanded zero-mode is forbidden because of $\mathbb{Z}_2$ symmetry in the $S^1/\mathbb{Z}_2$ orbifolding.
To generate the active neutrino masses, we have to introduce another righthanded neutrino $N$, which is not allowed to propagate in the extra dimension. This righthanded state couples to the active
states via tiny Yukawa couplings $y_0$, whereas the extra dimensional gauge singlet $\Psi_L^0$ couples to $N$ via dimensionful couplings $\kappa$. This way $\nu_L$ gets indirectly coupled to
the extra dimensional $\Psi_L^0$. The corresponding, CP conserving action yields
\begin{align}
\label{eq:lagrangian}
 S = \int \mathrm{d}^4x 
\begin{pmatrix}
 \overline{\nu_L} \, , & \overline{\Psi_L^0} \, , & \overline{N}
\end{pmatrix}
\begin{pmatrix}
 i\slashed \partial & 0 & y_0 v \\
 0 & i \slashed \partial + i \tilde{I}_{00} \partial_k \gamma^k & \kappa \\
 y_0 v & \kappa & i \slashed \partial
\end{pmatrix}
\begin{pmatrix}
   \nu_L \\ \Psi_L^0 \\ N
\end{pmatrix} \, ,
\end{align}
where $\tilde{I}_{00}$ is the mode diagonal correction parameter for the zero mode.
This parameter can be calculated analytically for the metric $\mathrm{d}s^2 = \mathrm{d}t^2 + \exp{(2kr\phi)} \mathrm{d}\vec{x}^2 + r^2 \mathrm{d} \phi^2 $ chosen here.
We obtain
\begin{align}
\label{eq:analyticCorr}
 \tilde{I}_{00} = \frac{1-\exp{(-4\pi kr)}}{4\pi kr} \, ,
\end{align}
in the case where the fundamental Dirac mass $m$ is much smaller than the warping scale parameter $k$. These parameters have to be chosen this way, since we want the left- and righthanded correction
integrals to be approximately equal to one another. \\
The effective vertex $\sim \overline{\Psi_L^0} \kappa N + h.c.$ can be achieved 
for example 
via a Higgs interaction by charging the righthanded $N$ under the same new symmetry group as the Higgs. This charge then naturally explains why $N$ cannot enter the extra 
dimension but is confined to the brane instead. Of course in this case the particle content must be further extended to make sure that gauge anomalies are cancelled out.  To discuss the dispersion relations, 
we perform a rotation from the interaction basis $\begin{pmatrix} \nu_L \,, &  \Psi_L^0 \, ,& N \end{pmatrix}$ to the propagation basis 
$\begin{pmatrix} \phi \,, &  \chi \, ,& \xi \end{pmatrix}$. 
For the propagations eigenstates $\phi$, $\chi$ and $\xi$ the dispersion relations can be calculated by variation $\delta S=0$ of the action, leading to the Euler-Lagrange equations for this particular problem 
(see Appendix C). 
The solutions for these ADR are given by
\begin{align}
  \label{eq:decDR}
 E^2_\phi &=  \vec{p}^2 \, , \\
 \label{eq:ADR}
 E^2_{\nicefrac{\chi}{\xi}} &= \kappa^2 + \vec{p}^2 \underbrace{\left[ \left(1+\frac{\tilde{I}_{00}}{2}\right)^2 -\frac{\tilde{I}_{00}^2}{4} \right]}_{f(\tilde{I}_{00})} 
\end{align}
where $\vec{p}$ is the 3-momentum on the brane. As expected, the decoupled dispersion relation \eqref{eq:decDR} does not get affected at zeroth order approximation for $y_0 v$, while the relations for the other two
eigenstates
are altered by a factor of $f(\tilde{I}_{00})$.
This can be interpreted as an altered dispersion relation allowing for sterile neutrino shortcuts in the 
extra dimension as suggested in the phenomenological approach of \cite{PaesShortcut}.
%This ADR can also be written in a more evocative way, setting the speed of light from $c=1$ back to $c$
%An effect caused by an ADR, which stems from asymmetrically warped metrics is a change in the probability function in neutrino oscillations, as already examined in
Rearranging Eq. \eqref{eq:ADR} and taking into account that $\tilde{I}_{00}\ll 1$ holds, we can expand the ADR in a Taylor series up to first order in $\tilde{I}_{00}$ additionally to the high energy limit $E \gg \kappa $,
yielding
\begin{align}
 p_{\nicefrac{\chi}{\xi}} \approx E - \frac{\kappa^2}{2E} - \frac{E^2}{2E} \tilde{I}_{00} + \mathcal{O}(\tilde{I}_{00}^2)+ \mathcal{O}(\kappa^2 \tilde{I}_{00})
\end{align}
for two of the propagation eigenstates $\chi$ and $\xi$. The contributions to the mass-squared-difference of the order $\mathcal{O}(\kappa^2 \tilde{I}_{00})$ are neglected, since it is only a renormalization of the 
coupling $\kappa$. Therefore we effectively end up with an additional, energy dependent potential $V_+ = E^2 \tilde{I}_{00}$, which induces new resonance 
phenomena in neutrino oscillations. Just as in \cite{PaesShortcut}, the potential $V_+$ has some properties, which are different from the standard matter potential induced from elastic forward scattering of active 
neutrinos and matter, being non-discriminatory between neutrinos and anti neutrinos and possessing a stronger energy dependence $V_+ \sim E^2$ instead of a linear dependence. 
The correction term $\tilde{I}_{00}$ and the shortcut parameter $\epsilon$ proposed in \cite{PaesShortcut} can be identified, when the mixing between active and sterile states is small. 
In this context we cannot confirm a hint towards baseline dependence of the resonant behaviour as suggested in \cite{PaesHollenbergBaseline} and consider it to
be an artifact of the semi-classical approach adopted in that work. 
In order to account for the correct resonance energy
while neglecting the effect of heavy KK excitations, one needs to vary the warp factor $k$ and extra dimensional radius $r$ independently. In order to explain the hierarchy problem one might be forced to
invoke more than one extra dimension.

\section{Conclusion}
In this paper we have developed a neutrino mass model giving rise to sterile neutrino
shortcuts in an asymmetrically warped extra dimension.  

In this context we have derived
the shape of the KK tower of an additional fermionic singlet in a general extra dimensional asymmetric warping framework and have demonstrated that this shape does not differ from symmetric warping scenarios (where the
warp factors of time and 3-space are the same). The main difference between both warping scenarios is
the emergence of effective
 Lorentz-violation on the 3-brane and a resulting altered dispersion relation of the fermionic singlet and
any particles mixing with it. 
 
Moreover, we have developed a concrete mechanism of neutrino mass generation, based on an additional particle
content consisting of a gauge singlet neutrino $\Psi$, which is able to propagate in the extra dimension, and an 
SM singlet neutrino $N$ confined to the 3-brane. 
The SM singlet brane neutrino mixes with the gauge singlet $\Psi$ and
the active neutrinos $\nu_L$ which conveys the effects of asymmetric warping to the active neutrinos, whose masses are generated by Yukawa interaction with the standard Higgs field. We have shown that such a 
model features an effective potential $V_+ \sim E^2$ during propagation, leading to resonant active-sterile neutrino oscillations. 

The sterile neutrino shortcut scenario proposed in \cite{PaesShortcut,Aeikens:2016rep,Aeikens:2014yga,SuperluminalNeutrinos,Doring:2018cob} can thus be understood as the four-dimensional effective field theory
limit of the model presented here.

\section{Acknowledgements}
HP thanks Andre de Gouvea for constant encouragement. This work was supported by 
DFG Grant No. PA 803/10-1.

\renewcommand{\bibname}{Bibliography}
\addcontentsline{toc}{chapter}{\bibname}
\bibliographystyle{unsrtdin}
\bibliography{AsymmetricWarpingBib}

\begin{thebibliography}{26}

% this bibliography is generated by unsrtdin.bst [8.2] from 2005-12-21

\providecommand{\url}[1]{\texttt{#1}}
\expandafter\ifx\csname urlstyle\endcsname\relax
  \providecommand{\doi}[1]{doi: #1}\else
  \providecommand{\doi}{doi: \begingroup \urlstyle{rm}\Url}\fi

\bibitem[1]{PaesShortcut}
\textsc{Paes}, Heinrich ; \textsc{Pakvasa}, Sandip  ; \textsc{Weiler},
  Thomas~J.:
\newblock {Sterile-active neutrino oscillations and shortcuts in the extra
  dimension}.
\newblock {In: }\emph{Phys. Rev.} D72 (2005), S. 095017.
\newblock \url{http://dx.doi.org/10.1103/PhysRevD.72.095017}. --
\newblock DOI 10.1103/PhysRevD.72.095017

\bibitem[2]{PaesHollenbergBaseline}
\textsc{Hollenberg}, Sebastian ; \textsc{Micu}, Octavian ; \textsc{Pas},
  Heinrich  ; \textsc{Weiler}, Thomas~J.:
\newblock {Baseline-dependent neutrino oscillations with extra-dimensional
  shortcuts}.
\newblock {In: }\emph{Phys. Rev.} D80 (2009), S. 093005.
\newblock \url{http://dx.doi.org/10.1103/PhysRevD.80.093005}. --
\newblock DOI 10.1103/PhysRevD.80.093005

\bibitem[3]{Aeikens:2016rep}
\textsc{Aeikens}, Elke ; \textsc{Päs}, Heinrich ; \textsc{Pakvasa}, Sandip  ;
  \textsc{Weiler}, Thomas~J.:
\newblock {Suppression of cosmological sterile neutrino production by altered
  dispersion relations}.
\newblock {In: }\emph{Phys. Rev.} D94 (2016), Nr. 11, S. 113010.
\newblock \url{http://dx.doi.org/10.1103/PhysRevD.94.113010}. --
\newblock DOI 10.1103/PhysRevD.94.113010

\bibitem[4]{Aeikens:2014yga}
\textsc{Aeikens}, Elke ; \textsc{Päs}, Heinrich ; \textsc{Pakvasa}, Sandip  ;
  \textsc{Sicking}, Philipp:
\newblock {Flavor ratios of extragalactic neutrinos and neutrino shortcuts in
  extra dimensions}.
\newblock {In: }\emph{JCAP} 1510 (2015), Nr. 10, S. 005.
\newblock \url{http://dx.doi.org/10.1088/1475-7516/2015/10/005}. --
\newblock DOI 10.1088/1475--7516/2015/10/005

\bibitem[5]{SuperluminalNeutrinos}
\textsc{Marfatia}, D. ; \textsc{Pas}, H. ; \textsc{Pakvasa}, S.  ;
  \textsc{Weiler}, T.~J.:
\newblock {A model of superluminal neutrinos}.
\newblock {In: }\emph{Phys. Lett.} B707 (2012), S. 553--557.
\newblock \url{http://dx.doi.org/10.1016/j.physletb.2012.01.028}. --
\newblock DOI 10.1016/j.physletb.2012.01.028

\bibitem[6]{Doring:2018cob}
\textsc{Döring}, Dominik ; \textsc{Päs}, Heinrich ; \textsc{Sicking}, Philipp
   ; \textsc{Weiler}, Thomas~J.:
\newblock {Sterile Neutrinos with Altered Dispersion Relations as an
  Explanation for the MiniBooNE, LSND, Gallium and Reactor Anomalies}.
\newblock   (2018)

\bibitem[7]{ADD}
\textsc{Arkani-Hamed}, Nima ; \textsc{Dimopoulos}, Savas  ; \textsc{Dvali},
  G.~R.:
\newblock {The Hierarchy problem and new dimensions at a millimeter}.
\newblock {In: }\emph{Phys. Lett.} B429 (1998), S. 263--272.
\newblock \url{http://dx.doi.org/10.1016/S0370-2693(98)00466-3}. --
\newblock DOI 10.1016/S0370--2693(98)00466--3

\bibitem[8]{RS}
\textsc{Randall}, Lisa ; \textsc{Sundrum}, Raman:
\newblock {A Large mass hierarchy from a small extra dimension}.
\newblock {In: }\emph{Phys. Rev. Lett.} 83 (1999), S. 3370--3373.
\newblock \url{http://dx.doi.org/10.1103/PhysRevLett.83.3370}. --
\newblock DOI 10.1103/PhysRevLett.83.3370

\bibitem[9]{Csaki:2000dm}
\textsc{Csaki}, Csaba ; \textsc{Erlich}, Joshua  ; \textsc{Grojean},
  Christophe:
\newblock {Gravitational Lorentz violations and adjustment of the cosmological
  constant in asymmetrically warped space-times}.
\newblock {In: }\emph{Nucl. Phys.} B604 (2001), S. 312--342.
\newblock \url{http://dx.doi.org/10.1016/S0550-3213(01)00175-4}. --
\newblock DOI 10.1016/S0550--3213(01)00175--4

\bibitem[10]{Chung:1999xg}
\textsc{Chung}, Daniel J.~H. ; \textsc{Freese}, Katherine:
\newblock {Can geodesics in extra dimensions solve the cosmological horizon
  problem?}
\newblock {In: }\emph{Phys. Rev.} D62 (2000), S. 063513.
\newblock \url{http://dx.doi.org/10.1103/PhysRevD.62.063513}. --
\newblock DOI 10.1103/PhysRevD.62.063513

\bibitem[11]{Chung:1999zs}
\textsc{Chung}, Daniel J.~H. ; \textsc{Freese}, Katherine:
\newblock {Cosmological challenges in theories with extra dimensions and
  remarks on the horizon problem}.
\newblock {In: }\emph{Phys. Rev.} D61 (2000), S. 023511.
\newblock \url{http://dx.doi.org/10.1103/PhysRevD.61.023511}. --
\newblock DOI 10.1103/PhysRevD.61.023511

\bibitem[12]{LSND1}
\textsc{Athanassopoulos}, C. u.\,a.:
\newblock Candidate Events in a Search for
  ${{\overline{\ensuremath{\nu}}}_{\ensuremath{\mu}}\ensuremath{\rightarrow}\overline{\ensuremath{\nu}}}_{\mathit{e}}$
  Oscillations.
\newblock {In: }\emph{Phys. Rev. Lett.} 75 (1995), Oct, 2650--2653.
\newblock \url{http://dx.doi.org/10.1103/PhysRevLett.75.2650}. --
\newblock DOI 10.1103/PhysRevLett.75.2650

\bibitem[13]{LSND2}
\textsc{Athanassopoulos}, C. u.\,a.:
\newblock Evidence for
  ${\overline{\ensuremath{\nu}}}_{\ensuremath{\mu}}\ensuremath{\rightarrow}{\overline{\ensuremath{\nu}}}_{\mathit{e}}$
  Oscillations from the LSND Experiment at the Los Alamos Meson Physics
  Facility.
\newblock {In: }\emph{Phys. Rev. Lett.} 77 (1996), Oct, 3082--3085.
\newblock \url{http://dx.doi.org/10.1103/PhysRevLett.77.3082}. --
\newblock DOI 10.1103/PhysRevLett.77.3082

\bibitem[14]{LSND3}
\textsc{Athanassopoulos}, C. u.\,a.:
\newblock Results on
  ${\ensuremath{\nu}}_{\mathit{\ensuremath{\mu}}}\phantom{\rule{0ex}{0ex}}\ensuremath{\rightarrow}\phantom{\rule{0ex}{0ex}}{\ensuremath{\nu}}_{\mathit{e}}$
  Neutrino Oscillations from the LSND Experiment.
\newblock {In: }\emph{Phys. Rev. Lett.} 81 (1998), Aug, 1774--1777.
\newblock \url{http://dx.doi.org/10.1103/PhysRevLett.81.1774}. --
\newblock DOI 10.1103/PhysRevLett.81.1774

\bibitem[15]{LSND4}
\textsc{Athanassopoulos}, C. u.\,a.:
\newblock Results on
  ${\ensuremath{\nu}}_{\ensuremath{\mu}}\ensuremath{\rightarrow}{\ensuremath{\nu}}_{e}$
  oscillations from pion decay in flight neutrinos.
\newblock {In: }\emph{Phys. Rev. C} 58 (1998), Oct, 2489--2511.
\newblock \url{http://dx.doi.org/10.1103/PhysRevC.58.2489}. --
\newblock DOI 10.1103/PhysRevC.58.2489

\bibitem[16]{LSND5}
\textsc{Aguilar-Arevalo}, A. u.\,a.:
\newblock {Evidence for neutrino oscillations from the observation of
  anti-neutrino(electron) appearance in a anti-neutrino(muon) beam}.
\newblock {In: }\emph{Phys. Rev.} D64 (2001), S. 112007.
\newblock \url{http://dx.doi.org/10.1103/PhysRevD.64.112007}. --
\newblock DOI 10.1103/PhysRevD.64.112007

\bibitem[17]{MiniBooNECombined}
\textsc{Aguilar-Arevalo}, A.~A. u.\,a.:
\newblock {A Combined $\nu_\mu \rightarrow \nu_e$ and $\bar \nu_\mu \rightarrow
  \bar \nu_e$ Oscillation Analysis of the MiniBooNE Excesses}, 2012

\bibitem[18]{MiniBooNE2}
\textsc{Aguilar-Arevalo}, A. A.; et~a.:
\newblock Improved Search for
  ${\overline{\ensuremath{\nu}}}_{\ensuremath{\mu}}\ensuremath{\rightarrow}{\overline{\ensuremath{\nu}}}_{e}$
  Oscillations in the MiniBooNE Experiment.
\newblock {In: }\emph{Phys. Rev. Lett.} 110 (2013), Apr, 161801.
\newblock \url{http://dx.doi.org/10.1103/PhysRevLett.110.161801}. --
\newblock DOI 10.1103/PhysRevLett.110.161801

\bibitem[19]{MiniBooNE2018}
\textsc{Aguilar-Arevalo}, A.~A. u.\,a.:
\newblock {Observation of a Significant Excess of Electron-Like Events in the
  MiniBooNE Short-Baseline Neutrino Experiment}.
\newblock   (2018)

\bibitem[20]{ThreeDetectorSBNO}
\textsc{Antonello}, M. u.\,a.:
\newblock {A Proposal for a Three Detector Short-Baseline Neutrino Oscillation
  Program in the Fermilab Booster Neutrino Beam}.
\newblock   (2015)

\bibitem[21]{TeraoMicroBooNE}
\textsc{Terao}, Kazuhiro:
\newblock {MicroBooNE: Liquid Argon TPC at Fermilab}.
\newblock {In: }\emph{JPS Conf. Proc.} 8 (2015), S. 023014.
\newblock \url{http://dx.doi.org/10.7566/JPSCP.8.023014}. --
\newblock DOI 10.7566/JPSCP.8.023014

\bibitem[22]{NeubertGrossman}
\textsc{Grossman}, Yuval ; \textsc{Neubert}, Matthias:
\newblock {Neutrino masses and mixings in nonfactorizable geometry}.
\newblock {In: }\emph{Phys. Lett.} B474 (2000), S. 361--371.
\newblock \url{http://dx.doi.org/10.1016/S0370-2693(00)00054-X}. --
\newblock DOI 10.1016/S0370--2693(00)00054--X

\bibitem[23]{KosteleckyShortBaseline}
\textsc{Kostelecky}, V.~A. ; \textsc{Mewes}, Matthew:
\newblock {Lorentz violation and short-baseline neutrino experiments}.
\newblock {In: }\emph{Phys. Rev.} D70 (2004), S. 076002.
\newblock \url{http://dx.doi.org/10.1103/PhysRevD.70.076002}. --
\newblock DOI 10.1103/PhysRevD.70.076002

\bibitem[24]{KosteleckyNeutrinos}
\textsc{Kostelecky}, Alan ; \textsc{Mewes}, Matthew:
\newblock {Neutrinos with Lorentz-violating operators of arbitrary dimension}.
\newblock {In: }\emph{Phys. Rev.} D85 (2012), S. 096005.
\newblock \url{http://dx.doi.org/10.1103/PhysRevD.85.096005}. --
\newblock DOI 10.1103/PhysRevD.85.096005

\bibitem[25]{Antonelli}
\textsc{Antonelli}, V. ; \textsc{Miramonti}, L.  ; \textsc{Torri}, M. D.~C.:
\newblock {Neutrino oscillations and Lorentz Invariance Violation in a
  Finslerian Geometrical model}.
\newblock   (2018)

\end{thebibliography}

\section{Appendix}
\subsection{A: Calculation of the KK spectrum}
From all the terms that are contained in Equation \eqref{eq:generalaction}, the ones containing greek indices are analyzed first.
This is because of their similarities regarding their spinor structure $\overline{\Psi}_\halb \mathcal{O}(\partial_\mu) \Psi_\halb$, 
making them comparable to a kinetic term. They are
\begin{align}
 \begin{split}
S \supseteq S_\text{kin} = \int \mathrm{d}^4 x \int \mathrm{d}\phi \sqrt{\text{det}G} \left\lbrace \vphantom{\frac{\sqrt{\text{det}G}}{2B}} \right.
&\frac{i}{2A} \left[ \overline{\Psi}_1 \gamma^0 \left( \partial_0 - \overleftarrow{\partial_0} \right) \Psi_1 
+ \overline{\Psi}_2 \gamma^0 \left( \partial_0 - \overleftarrow{\partial_0} \right) \Psi_2 \right]  \\
+&\frac{i}{2B} \left[ \overline{\Psi}_1 \gamma^k \left( \partial_k - \overleftarrow{\partial_k} \right) \Psi_1 
+ \overline{\Psi}_2 \gamma^k \left( \partial_k - \overleftarrow{\partial_k} \right) \Psi_2 \right] \left. \vphantom{\frac{1}{2C}}
\right\rbrace \, .
 \end{split}
\end{align}
To convert the left-bound derivatives into standard (right-bound) derivatives, an integration-by-parts is used:
\begin{align}
 \begin{split}
S_\text{kin} = \int \mathrm{d}^4 x \int \mathrm{d}\phi \, i  \left\lbrace 
+\overline{\Psi}_1 \gamma^0 \frac{\sqrt{\text{det}G}}{2A} \,  \partial_0 \Psi_1 
+\overline{\Psi}_2 \gamma^0 \frac{\sqrt{\text{det}G}}{2A} \,  \partial_0 \Psi_2 \right.  \\ \allowbreak
+\overline{\Psi}_1 \gamma^k \frac{\sqrt{\text{det}G}}{2B} \,  \partial_k \Psi_1 
+\overline{\Psi}_2 \gamma^k \frac{\sqrt{\text{det}G}}{2B} \,  \partial_k \Psi_2 \left.\vphantom{\frac{1}{2B}} \right\rbrace  \\
-\int \mathrm{d}^3 x^k \int \mathrm{d}\phi \, i 
\left\lbrace \underbrace{\left[ \overline{\Psi}_1 \gamma^0 \frac{\sqrt{\text{det}G}}{2A} \, \Psi_1 \right]_{\partial V}}_{=0}
-\underbrace{\left[ \overline{\Psi}_2 \gamma^0 \frac{\sqrt{\text{det}G}}{2A} \, \Psi_2 \right]_{\partial V}}_{=0} \right\rbrace \\
-\int \mathrm{d}^{4-k} x \int \mathrm{d}\phi \, i
\left\lbrace \underbrace{\left[ \overline{\Psi}_1 \gamma^k \frac{\sqrt{\text{det}G}}{2B} \, \Psi_1 \right]_{\partial V}}_{=0}
-\underbrace{\left[ \overline{\Psi}_2 \gamma^k \frac{\sqrt{\text{det}G}}{2B} \, \Psi_2 \right]_{\partial V}}_{=0} \right\rbrace \\
+ \int \mathrm{d}\phi \, i  \left\lbrace 
\overline{\Psi}_1 \gamma^0  \partial_0 \frac{\sqrt{\text{det}G}}{2A} \,  \Psi_1 
+\overline{\Psi}_2 \gamma^0  \partial_0 \frac{\sqrt{\text{det}G}}{2A} \,  \Psi_2 \right. 
\\ \left.
+\overline{\Psi}_1 \gamma^k  \partial_k \frac{\sqrt{\text{det}G}}{2B} \,  \Psi_1 
+\overline{\Psi}_2 \gamma^k  \partial_k \frac{\sqrt{\text{det}G}}{2B} \,  \Psi_2 \right\rbrace  \, .
 \end{split}
\end{align}
The terms on the border vanish since it is assumed that all quantum fields vanish in infinity.
Since the entries of the metric $G$ are not dependent on the brane coordinates, the derivatives commute with the operator expression 
$\partial_\mu \nicefrac{ \sqrt{\text{det}G(\phi)}}{2f(\phi)}= \nicefrac{ \sqrt{\text{det}G(\phi)}}{2f(\phi)} \, \partial_\mu $
and the action can be written as in Eq. \eqref{eq:kineticaction}. This form allows for a 'smooth' KK decomposition. \\
In the second analysis of the five dimensional action $S$, the terms containing derivatives with respect to the extra dimension $\partial_\phi$ 
are under examination. Due to the configuration of their spinors, these terms are connected to the mass-term. The relevant terms are
 \begin{align}
 \begin{split}
S \supseteq S_\text{mass} = \int \mathrm{d}^4 x \int \mathrm{d}\phi \sqrt{\text{det}G} \left\lbrace \vphantom{\frac{1}{2}} \right. 
&-\frac{1}{2rC} \left[ \overline{\Psi}_1 \gamma^5 \left( \partial_\phi - \overleftarrow{\partial_\phi} \right) \Psi_2 
+\overline{\Psi}_1 \gamma^5 \left( \partial_\phi - \overleftarrow{\partial_\phi} \right) \Psi_2 \right]  \\
&-m \, \text{sgn}(\phi) \left[ \, \overline{\Psi}_1 \Psi_2 + \overline{\Psi}_2 \Psi_1 \right] \left. \vphantom{\frac{1}{2C}} \right\rbrace \, .
 \end{split}
\end{align}
In analogy to the calculations in the terms with indices $\mu$, an integration-by-parts is conducted and using the relationship
\begin{align}
 \overline{\Psi}_\halb \gamma^5 \Psi_{\nicefrac{2}{1}} = \pm \overline{\Psi}_\halb \Psi_{\nicefrac{2}{1}} 
\end{align}
we arrive at Eq. \eqref{eq:massaction}.
which can be decomposed via Eq. \eqref{eq:ansatz}.
The kinetic part after decomposition reads
\begin{align}
 \label{eq:kineticDecomposed}
 \begin{split}
S_{\text{kin}} = \int \mathrm{d}^4 x \int \mathrm{d}\phi \sum_n \sum_m  \left\lbrace \vphantom{\frac{\gamma^0}{A}} \frac{C}{A} \right. & \left[ 
\overline{\psi^1_n} \hat{f}^{1 \dagger}_n \, i 
\left( \slashed \partial + \frac{(A-B)\gamma^k}{B} \partial_k \right)  \psi_m^1  \hat{f}_m^1 \right. 
\\
&\left. \left.
+\overline{\psi^2_n} \hat{f}^{2 \dagger}_n \, i 
\left( \slashed \partial + \frac{(A-B)\gamma^k}{B} \partial_k \right) \psi_m^2  \hat{f}_m^2 
\right] \right\rbrace \, ,
 \end{split}
\end{align}
from which we can infer the scalar product in Eq. \eqref{eq:scalarproduct} and the correction term \eqref{eq:OmegaTerm} by matching it to the corrected Dirac action \eqref{eq:correctedDirac}. \\
With these conditions set, we decompose the mass term of the action and find
\begin{align}
 \begin{split}
S_{\text{mass}} &= \int \mathrm{d}^4 x \int \mathrm{d}\phi \sum_n \sum_m  \left\lbrace - \left[ 
\overline{\psi^1_n}\frac{1}{\sqrt{2r\xi}} \hat{f}^{1 \dagger}_n 
\left( \xi  \, \partial_\phi + \partial_\phi \xi  \, \right)
\psi_m^2 \frac{1}{\sqrt{2r\xi}} \hat{f}_m^2 \right. \right. \\
& \hspace{4.7cm} \left. - \overline{\psi^2_n}\frac{1}{\sqrt{2r\xi}} \hat{f}^{2 \dagger}_n  
\left( \xi \, \partial_\phi + \partial_\phi\xi  \, \right)
\psi_m^1 \frac{1}{\sqrt{2r\xi}} \hat{f}_m^1 \right] \\
&\left.- 2rC \xi \,m \, \frac{\text{sgn}(\phi)}{2r\xi} \left[ \, 
 \overline{\psi^1_n} \hat{f}^{1 \dagger}_n \cdot \psi_m^2 \hat{f}_m^2  
+\overline{\psi^2_n} \hat{f}^{2 \dagger}_n \cdot \psi_m^1 \hat{f}_m^1 \right] 
\vphantom{\frac{\sqrt{\text{det}G}}{2rC}} \right\rbrace  \, .
 \end{split}
\end{align}
Carrying out derivatives leads to terms, which cancel in a nice way because of the carefully chosen ansatz \eqref{eq:ansatz} and we obtain
\begin{align}
\label{eq:stower}
 \begin{split}
S_{\text{mass}} = \int \mathrm{d}^4 x \int \mathrm{d}\phi \sum_n \sum_m  &\left\lbrace 
\overline{\psi_n^1} \hat{f}_n^{1 \dagger} \left(- \frac{\partial_\phi}{r} - m \, \text{sgn}(\phi) C \right) \psi_m^2 \hat{f}_m^2 \right.
\\
&\left. +\overline{\psi_n^2} \hat{f}_n^{2 \dagger} \left(+ \frac{\partial_\phi}{r} - m \, \text{sgn}(\phi) C \right) \psi_m^1 \hat{f}_m^1
\right\rbrace \, .
 \end{split}
\end{align}
To make use of the already determined scalar product \eqref{eq:scalarproduct}, the functions $\hat{f}_n^\halb$ are constructed as eigenfunctions
of the hermitian operator $\left( \pm \frac{\partial_\phi}{r} - m C' \right)$. This condition is a system of coupled, first order differential equations \eqref{eq:dglasymm}, which fixes the shape of the KK spectrum,
while Eq. \eqref{eq:scalarproduct} fixes normalization.
\subsection{B: Calculation of the Zero Mode Correction Integral}
Beginning with Eq. \eqref{eq:dglasymm}, we make the zeroth KK mass $M_0$ to vanish. This way, the equations decouple and we are left with a simple euqation
\begin{align}
 \left(\mp \frac{\partial_\phi}{r} - m\right) \hat{f}^{\nicefrac{2}{1}}_0 = 0 \, .
\end{align}
This equation can be solved via the separation of variables method, yielding
\begin{align}
 \hat{f}^{\nicefrac{2}{1}}_0(\phi) = \hat{K}^{\nicefrac{2}{1}}_0 \exp{(\mp 2rm \phi)} \, .
\end{align}
The second condition in Eq. \eqref{eq:scalarproduct}, the scalar product of the defined functions, gives us the normalization
\begin{align}
 \int \mathrm{d}\phi \hat{f}^{\nicefrac{2}{1}}_0(\phi) \hat{f}^{\nicefrac{2}{1} \, \dagger}_0(\phi) = 1 \\
 \Rightarrow \left(\hat{K}^{\nicefrac{2}{1}}_0\right)^2 = \frac{\mp 2rm}{\exp{(\mp 4\pi rm)}-1}
\end{align}
Using the definition of the correction integral 
$\tilde{I}^{\nicefrac{2}{1}}_{00} = \int \mathrm{d}\phi \hat{f}^{\nicefrac{2}{1} \, \dagger}_0 \frac{C}{B} \hat{f}^{\nicefrac{2}{1}}_0 -1$
, we obtain
\begin{align}
 \tilde{I}^{\nicefrac{2}{1}}_{00} = \frac{\frac{m}{k}}{\frac{m}{k} \pm 1} \frac{\exp{\left( 4\pi kr \left( \mp \frac{m}{k} -1\right)  \right)}-1}{\exp{\left(\mp 4\pi kr \frac{m}{k}  \right)}-1}\, .
\end{align}
In the limit of $\frac{m}{k} \to 0$ an equality between $\tilde{I}^{1}_{00}$ and $\tilde{I}^{2}_{00}$ is achieved.
In this case the correction integral breaks down to the value in Eq. \eqref{eq:analyticCorr}.

\subsection{C: Calculation of the ADR}
Starting from the 4D-Lagrangian $\mathcal{L}_\text{int} = \overline{v} \uuline{\mathcal{L}_\text{int}} v $ in Eq. \eqref{eq:lagrangian} in interaction space, we rotate to propagation space via a unitary transformation
$U$ so that
\begin{align}
 \mathcal{L}_\text{int} = 
\begin{pmatrix}
 \overline{\phi} \, , & \overline{\chi} \, , & \overline{\xi}
\end{pmatrix}
\underbrace{U^\dagger \uuline{\mathcal{L}_\text{int}} \,  U}_{=: \uuline{\mathcal{L}_\text{pro}}}
\begin{pmatrix}
   \phi \\ \chi \\ \xi
\end{pmatrix} 
\end{align}
holds. The eigenvalue equation 
\begin{align}
 \text{det}\left( \uuline{\mathcal{L}_\text{int}} - \mathds{1}_{3} \otimes \lambda \right) = 0
\end{align}
of the external structure can be written in Fourier space as 
\begin{align}
\begin{split}
\label{eq:determinant}
\det{\begin{pmatrix} 
        \mathbb{V} & \mathbb{W} \\
        \mathbb{X} & \mathbb{Y}
      \end{pmatrix}}
      &= \det{\left[ 
      \underbrace{\begin{pmatrix} 
        \mathbb{V} & 0 \\
        \mathbb{X} & \mathds{1}
      \end{pmatrix}}_{:=\mathbb{A}}
      \underbrace{\begin{pmatrix} 
        \mathds{1} & \mathbb{V}^{-1} \mathbb{W} \\
	0 & \mathbb{Y}-\mathbb{X}\mathbb{V}^{-1}\mathbb{W}
      \end{pmatrix}}_{:=\mathbb{B}}
      \right]} \hspace{1cm} \text{,if } \mathbb{V} \text{ is invertible} \\
      &= \det{(\mathbb{A})} \det{(\mathbb{B})} \\
      &= \det{(\mathbb{V})} \det{(\mathbb{Y}- \mathbb{X}\mathbb{V}^{-1}\mathbb{W})} \, .
  \end{split}
\end{align}
where 
\begin{align}
 \begin{split}
  \mathbb{V} &= \slashed p \hspace{4cm}
  \mathbb{W} = \begin{pmatrix} 0 & y_0 v \end{pmatrix} \\
  \mathbb{X} &= \begin{pmatrix} 0 & y_0 v \end{pmatrix}^T  \hspace{2.5cm}
  \mathbb{Y} = \begin{pmatrix} \slashed p + \tilde{I}_{00} p_k \gamma^k & \kappa \\ \kappa & \slashed p \end{pmatrix} \, .
 \end{split}
\end{align}
This leads to the solution
\begin{align}
 0 &= \left( \slashed p -\lambda \right)^3 + \tilde{I}_{00} p_k \gamma^k \left[ (\slashed p - \lambda)^2 - y_0^2 v^2 \right] - \left[ y_0^2 v^2 - \kappa^2  \right] (\slashed p -\lambda)  \\
  &\stackrel{y^2 \ll 1}{\Rightarrow} \lambda \approx \slashed p \,  \vee \,  \lambda \approx \slashed p + \frac{\tilde{I}_{00} p_k \gamma^k}{2} \pm \frac{\sqrt{4 \kappa^2 + \left( \tilde{I}_{00} p_k \gamma^k\right)^2}}{2} \, .
\end{align}
These correspond to the eigenvalues of propagation. \\

\begin{comment}
To determine the dispersion relation, we use the Euler-Lagrange equation for the inner structure of propagation eigenstates and find
\begin{align}
  \begin{split}
   \frac{\partial \mathcal{L}}{\partial \overline{\chi}} = \left( \slashed p + \frac{\tilde{I}_{00} p_k \gamma^k}{2} + \frac{\sqrt{4 \kappa^2 + \left( \tilde{I}_{00} p_k \gamma^k\right)^2}}{2} \right) \chi &= 0 \\
   \frac{\partial \mathcal{L}}{\partial \overline{\xi }} = \left( \slashed p + \frac{\tilde{I}_{00} p_k \gamma^k}{2} - \frac{\sqrt{4 \kappa^2 + \left( \tilde{I}_{00} p_k \gamma^k\right)^2}}{2} \right) \xi &= 0 \, .
  \end{split}
\end{align}
To find the conditions for this to be solved, we again have to calculate the determinant
\begin{align}
 \text{det} 
 \begin{pmatrix}
  E\pm \frac{\sqrt{4\kappa^2 - \vec{p}^2 \tilde{I}_{00} } }{2} & p_k \sigma^k \left( 1+ \frac{\tilde{I}_{00}}{2} \right) \\
  -p_k \sigma^k \left( 1+ \frac{\tilde{I}_{00}}{2} \right) & -E\pm \frac{\sqrt{4\kappa^2 - \vec{p}^2 \tilde{I}_{00} } }{2}
 \end{pmatrix}
=0
\end{align}
and use the decomposition in \eqref{eq:determinant}. The sign $\pm$ depends on the choice of eigenstate ($\chi\to +$, $\xi \to -$), but is unimportant to the result. For the respresentation of the $\gamma$-matrices we
use the Dirac respresentation in 4D, since the extra dimension is already integrated out. The solution of this equation is the dispersion relation in Eq. \eqref{eq:ADR}. 
\end{comment}
\end{document}